\documentclass{ESLAB2005}
\usepackage{epsfig}
\begin{document}
\ifx\href\undefined\else\hypersetup{linktocpage=true}\fi 

\title{Opening a New Window to Fundamental Physics and Astrophysics: X-ray Polarimetry}

\author{E. Costa\inst{1}, R. Bellazzini\inst{2}, P. Soffitta\inst{1},
G. di Persio\inst{1}, M. Feroci\inst{1}, E. Morelli\inst{1}, F.
Muleri\inst{1}, L. Pacciani\inst{1}, A. Rubini\inst{1}, L.
Baldini\inst{2}, F. Bitti\inst{2}, A. Brez\inst{2}, F.
Cavalca\inst{2}, L. Latronico\inst{2}, M. M. Massai\inst{2}, N.
Omodei\inst{2}, C. Sgro'\inst{2},  G. Spandre\inst{2}, G.
Matt\inst{3}, G. C. Perola\inst{3}, A. Santangelo\inst{4},  A.
Celotti\inst{5}, D. Barret\inst{6}, O. Vilhu\inst{7}, L.
Piro\inst{1}, G. Fraser\inst{8}, T. J.-L. Courvoisier\inst{9}, X.
Barcons\inst{10}} \institute{Istituto di Astrofisica Spaziale e
Fisica Cosmica of INAF, Via del Fosso del Cavaliere, 100, I-00133
Roma, Italy \and Sezione di Pisa of INFN, Largo B. Pontecorvo, 3
I-56127 Pisa, \and Dipartimento di Fisica -- Universita' di
Roma-3, via della Vasca Navale 84, I-00146 Roma, Italy, \and
Institut fuer Astronomie und Astrophysik der Eberhard Karls
Universitaet, Sand 1 - D-72076 Tuebingen, Geramy, \and S.I.S.S.A,
via Beirut 2-4, 34014 Trieste, Italy, \and Centre d'Etude Spatiale
des Rayonnements, 9, avenue du Colonel Roche, 31028 Toulouse Cedex
4, France\and Observatory -- University of Helsinki, box 14
fin-00014, Helsinki, Finland,\and Space Research Centre,
University of Leicester, University Road Leicester LE1 7RH United
Kingdom \and ISDC, 16, ch. d'Ecogia, CH-1290 VERSOIX, Switzerland,
\and Instituto de F\'{\i}sica de Cantabria (CSIC-UC), E-39005
Santander, Spain.}

\maketitle

\begin{abstract}

An extensive theoretical literature predicts that X-ray
Polarimetry can directly determine relevant physical and
geometrical parameters of astrophysical sources, and discriminate
between models further than allowed by spectral and timing data
only. X-ray Polarimetry can also provide tests of Fundamental
Physics.

A high sensitivity polarimeter in the focal plane of a New
Generation X-ray telescope could open this new window in the High
Energy Sky.

\end{abstract}

\section{Introduction}

Since the birth of X-ray astronomy, spatial, spectral and timing
observations have improved dramatically in quantity and quality,
providing a wealth of information on almost all classes of cosmic
objects. One key aspect of the emission, however, remained
basically unprobed: polarization. Conventional polarimeters are
cumbersome and poorly sensitive, so that they were not included in
the major missions.

The lack of polarimetric measurements is extremely painful, as
they would add two essential parameters (amount and angle of
polarization) to those already derived from spectra and timing,
providing unique  information on the Geometry and Physics of
sources belonging to most of classes of interest (\cite{mesz88}).

The advent of new generation, focal plane, photoelectric
polarimeters (\cite{costa01}, \cite{bellazzini03}), all developed
in Europe, provides the breakthrough in sensitivity needed to
search for polarization degrees of astrophysical interest (a few
percent) in several classes of cosmic sources. We discuss here
briefly a few scientific cases of outstanding interest.

\section{What can Polarimetry Test?}
\label{sec:cmd}

\subsection{Astrophysics}
\label{sec:ex} Polarization from celestial sources may derive
from:
\begin{itemize}
\item Emission processes themselves:    cyclotron, synchrotron,
non-thermal bremmstrahlung (\cite{west59}, \cite{gnesun},
\cite{rees75})
\item Scattering on aspheric accreting
plasmas: disks, blobs, columns. Resonant scattering of lines in
hot plasmas (\cite{rees75}, \cite{sun-tit}, \cite{mesz88},
\cite{saslines}).
\item Vacuum polarization and birefringence through extreme
magnetic fields (\cite{gneal78}; Ventura, 1979; \cite{mesz80})
\end{itemize}
All these processes produce relevant polarization that, unless
averaged and smeared by a symmetric distribution of the involved
regions, will be detectable in a variety of situations and classes
of sources:
\begin{itemize}
\item  ordered acceleration
\item  accretion
\item  reflection present or past (archaeoastronomy)
\item  jets

\end{itemize}

\subsection{Fundamental Physics}
\label{sec:ex}
\begin{itemize}
\item  Matter in extreme magnetic fields
\item  Matter in strong gravity fields
\item  Long Distance Quantum Gravity effects
\end{itemize}

\subsection{Big Hopes Meagre Results}
Notwithstanding these big expectations, polarimetry, one key
aspect of the emission, has remained basically unexplored. After
some pioneering rocket experiment, two polarimeters have been
flown aboard ARIEL-5 and OSO-8. So far, only one detection, the
Crab Nebula, (\cite{nov72}, \cite{weis76}, \cite{weis78}) and a
handful of not very significant upper limits (\cite{silver78},
\cite{hug84}) are available, all of them obtained with the Bragg
Polarimeter on-board OSO-8. The results of Polarimetry are indeed
marginal in X-ray Astronomy.
\subsection{The Technique is the limit}
The technique conventionally exploited to measure linear
polarization is bragg diffraction at $45^{\circ}$. The whole
instrument is rotated around the axis and the modulation of the
diffracted flux at each angle provides the amount of polarization.
Due to the narrow band of the process, there is a good control of
systematics and imaging is preserved, but, even with the best
mosaic crystals, the integrated efficiency is extremely low.

The other conventional technique is Compton scattering around
$90^{\circ}$. The technique is more effective, but only above 5
keV, is not imaging, heavily dominated by background and affected
by systematics.

Both techniques require rotation around the optical axis, are
relatively cumbersome and yield a low sensitivity.

\section{A new device: the Micropattern Gas Chamber}
\begin{figure}[ht]
\begin{center}
\epsfig{file=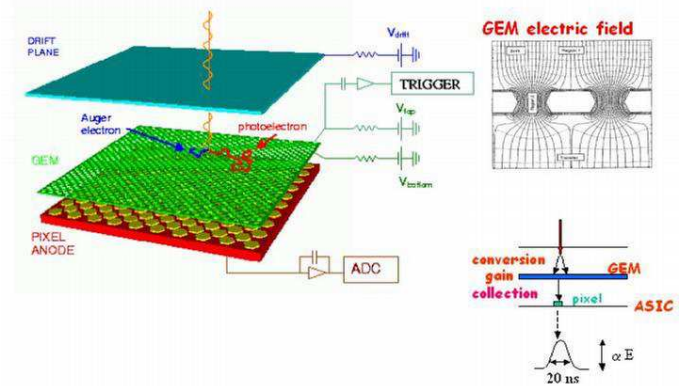, width=9cm}
\end{center}
\caption{The Concept of Micropattern Gas Chamber \label{MPGC}}
\end{figure}

\begin{figure}[ht]
\begin{center}
\epsfig{file=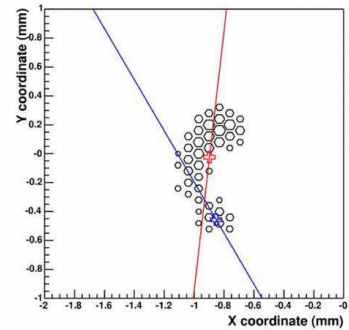, width=7cm}
\end{center}

\caption{The track produced in the gas by a 5.9 keV photon is
recorded by the MPGC and analyzed with the following sequence:
barycenter evaluation; reconstruction of the principal axis;
reconstruction of the conversion point; reconstruction of emission
direction. The polarization is derived from the reconstructed
angular distribution of the photoelectrons.\label{track}}
\end{figure}

A new device that analyzes linear polarization, while preserving
the information on the point of impact, the energy and the time,
is now available: the Micropattern Gas Chamber. It represented, at
the beginning, the extreme evolution of the imaging proportional
counter, with a planar multiplication stage and a multi-anode
read-out. Now it evolved into a full pixel detector, with a gas
converter.  The X-ray photon is absorbed in a low Z mixture,
producing a photoelectron that ionizes the gas. The result is a
ionization track. Electrons drift to a Gas Electron Multiplier,
are multiplied and collected from a read-out plane, made of metal
pads, each with its individual electronic chain (Fig.~\ref{MPGC}).
From the analysis of the track image (see Fig.~\ref{track}), the
interaction point and the photoelectron direction are
reconstructed. The angular distribution of the photoelectron
directions depends on linear polarization with a $cos^{2}$
distribution.

\begin{figure}[ht]
\begin{center}
\epsfig{file=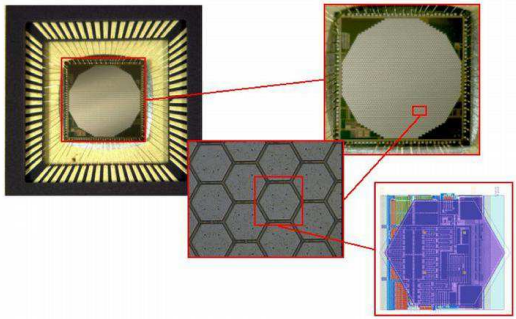, width=7cm}
\end{center}

\caption{Progressively zoomed photos of the Chip of generation I
(2100 pixel) showing the pad structure. The last zooming shows the
drawing of the read out electronics for a single pixel harbored in
silicon layers below the pad.} \label{zoom}
\end{figure}

\begin{figure}[ht]
\begin{center}
\epsfig{file=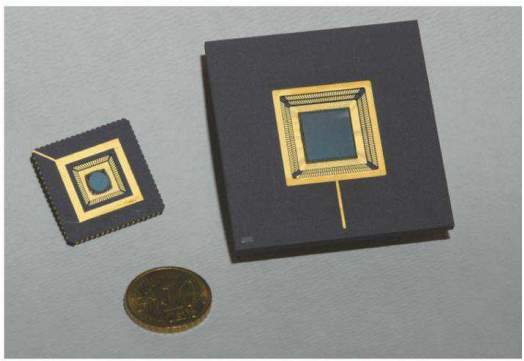, width=7cm}
\end{center}

\caption{VLSI read-out chips of generation I(2100)pixels and II
(20000 pixels) \label{chips}}
\end{figure}

In a first pioneering phase the read-out was based on a
multi-layer PCB routing the signals of each pad to an external
analysis chain, resident on an ASIC chip. In a further stage a
VLSI chip was developed including the metal pads acting as an
anode and a complete electronic chain under the projection of each
pad, in lower layers(Fig.~\ref{zoom}). On trigger from the GEM
signals are routed to an external ADC. The whole is extremely
compact and allowed for a fast development of chips of increased
dimension (Fig.~\ref{chips}) with a progressively decreasing pixel
size (see Table~\ref{tab:table}). The last generation chip arrived
to 100000 pixels of $50 \mu$m pitch. To reduce the read-out dead
time the new chip is capable to self trigger and fetch only a
window around the trigger pixel.
\begin{table}[bht]
  \caption{The evolution of the read-out from the first
  prototype, based on ML PCB technique, to the ASIC chips.}
  \label{tab:table}
  \begin{center}
    \leavevmode
    \footnotesize
    \begin{tabular}[h]{lrccc}
      \hline \\[-5pt]
      Device&Epoch& pixels&pixel($\mu$m)& size (mm) \\[+5pt]
      \hline \\[-5pt]
      Protype PCB&2001& 500 & 120 & 3 diam \\
      Chip I & 2003 & 2100 & 80 & 4 diam\\
      Chip II &2004 & 20000 & 80 & 11 x 11 \\
      Chp III &in prod.&100000& 50 & 17 x 17 \\
      \hline \\
      \end{tabular}
  \end{center}
\end{table}

\begin{figure}[ht]
\begin{center}
\epsfig{file=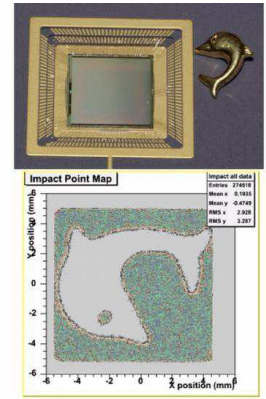, width=7cm}
\end{center}

\caption{Image of an ear ring done with a detector based on chip
of generation II and processed with the reconstruction of the
impact point\label{delfino}}
\end{figure}

This new device in the focal plane of a New Generation X-Ray
Telescope with an effective area $\geq 10 m^{2}$  can turn the
dream of X-ray Polarimetry into reality. The critical parts (GEM
and VLSI) do exist: body, window, gas handling, HV etc. are
conventional, well established technology. It is relatively small
and compact: no cryogenics, no rotations and the VLSI is radiation
hard. It is very fast: can handle high rates (5$\times 10^{4}$
c/s). It is position sensitive to around 100 $\mu$ m: oversamples
the p.s.f. of any optics but Chandra. While doing Polarimetry also
performs timing, imaging (Fig.~\ref{delfino}) and spectra (with
the energy resolution of a good PC).

We tested the device searching for systematic effects, namely
spurious polarization, when measuring photons from an unpolarized
(fluorescence) source. We do not find any effect down to 0.2\% ,
so that this is our present limit, to be possibly improved.
\section{Which Polarimetry with this new device?}
\subsection{The strange case of Sgr B2} 
\begin{figure}[ht]
\begin{center}
\epsfig{file=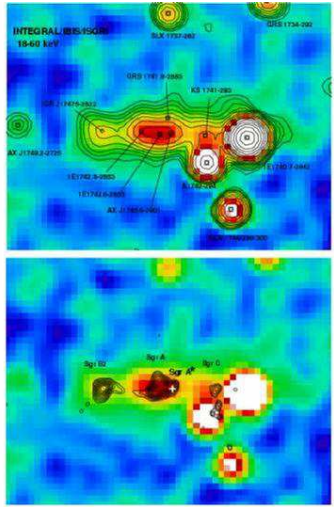, width=9cm}
\end{center}

\caption{The Galactic Center as imaged with INTEGRAL-IBIS
experiment} \label{GCIBIS}
\end{figure}

\begin{figure}[ht]
\begin{center}
\epsfig{file=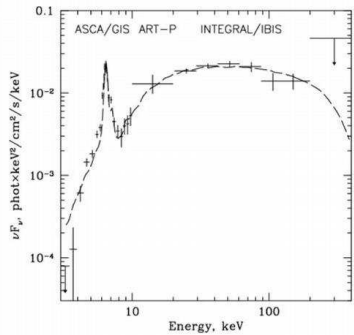, width=7cm}
\end{center}

\caption{The Spectrum of SgrB2 by combining data from ASCA and
IBIS is a pure reflection spectrum. The input spectrum corresponds
to a power law with photon index 1.8. } \label{GCSPECTR}
\end{figure}

Interaction of Supermassive
Black Holes and their host galaxy is a clue of High Energy
Astrophysics in the Cosmic Vision Era. Most of supermassive BH at
the center of galaxies are nowadays inactive. A few are very
active. How do they swap from one status to the other? A
noticeable case: our own Galaxy. The discovery of a
2.6$\times10^6$ solar masses Black Hole at the center of our own
Galaxy was certainly one of the most exciting result in Astronomy
in recent years. The Black Hole is very quiet, its accretion
luminosity being about 10 orders of magnitude lower than the
Eddington luminosity. However, if Brahe or Galileo could have at
their disposal a X-ray satellite, they would possibly have
observed the Galactic Center as the brightest extrasolar X-ray
source. In fact, at the projected distance of about 100 pc from
the Black Hole, there is a giant molecular cloud, Sgr
B2(Fig.~\ref{GCIBIS}), which in X-rays has a pure reflection
spectrum (Fig.~\ref{GCSPECTR}, \cite{koy96}). However, it is not
clear what Sgr B2 is reflecting: there are no bright enough
sources in the vicinity. The simplest explanation is that a few
hundreds years ago our own Galactic Center was much brighter, at
the level of a low luminosity Active Galaxy: the molecular cloud
would then simply echoing the past activity (\cite{sun93},
\cite{koy96}).
\begin{figure}[ht]
\begin{center}
\epsfig{file=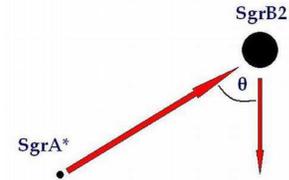, width=4cm}
\end{center}

\caption{We only observe a projected distance of 100pc from SgrA
and SgrB2. The amount of polarization will provide the angle
$\vartheta$ and, fix the distance of the two sources and thence
give the epoch when our GC was an AGN} \label{GCangle}
\end{figure}
Polarimetric measurements would be able to confirm or disproof
this hypothesis beyond doubts (\cite{chur02}): not only reflected
radiation should be highly polarized, but the polarization angle
must be perpendicular to the projected line connecting the Black
Hole to Sgr B2. The degree of polarization will also tell us the
true distance of Sgr B2 (Fig.~\ref{GCangle}), much tightly
constraining the time at which the Galactic Center was active.

\subsection{Matter under extreme magnetic fields } Magnetized
neutron stars are an excellent laboratory to test QED effects in
the presence of magnetic fields that cannot be produced in
laboratories. These span from the well established fields of
$10^{12} - 10^{13}$ gauss of accreting X-ray pulsators to the
extreme fields up to $10^{15}$ gauss proposed for Soft Gamma
Repeaters and Anomalous X-ray Pulsars.

In \textbf{isolated neutron stars}a soft thermal emission is
produced in hot spots at the surface of the star or close to it.
This is a detected as a black-body emission smoothly modulated by
the rotation of the star. The radiation to arrive to the observer
will cross a region of vacuum with strong magnetic field, whose
intensity and orientation will be phase dependent. But QED
predicts a significant birefringence of the vacuum that will not
affect the continuum energy spectrum, but will introduce a very
strong, phase dependent, polarization, allowing to map the
geometry and intensity of the magnetic field (\cite{pavzav00},
\cite{hesh02}, \cite{laiho03}). Isolated X-ray pulsars are too
faint and too soft to make this measurement feasible with the
present status of the MPGC technology, but could become realistic
with an improved detector. Anyhow the same phenomenology is
expected (and should be even more outstanding) in the case of Soft
Gamma Repeaters. The flux of SGRs, even in quiescent state, is
compatible with such a measurement. Due to the extreme magnetic
field also two absorption features should be present, one at the
so called vacuum resonance frequency, and one at the
proton-cyclotron resonance (\cite{nimbul05}). These features
should have different phenomenology with respect to polarization.
This would provide a direct evidence of the presence of extreme
magnetic fields and provide a check of the magnetar model, so far
mainly supported by energetic considerations. In an active phase
or in the first few days following a major emission episode the
situation may become even more interesting, because of the likely
presence of a transient ionized atmosphere. In this frame it would
be extremely attractive a measurement of the lines such as the one
detected at 5.0 keV in the spectrum of SGR1806-20 (\cite{ibr02}).
It has been suggested (\cite{zane01}) that, this feature are
proton cyclotron in the presence of a magnetic field of 1.0
$\times 10^{15}$ gauss. This completely exotic scenario can be
deeply probed with polarimetry. In fact the energy resolved
evolution of the polarization amount and angle following a burst
episode will provide a true laboratory to test QED prediction and
to investigate the nature of the delayed emission of magnetar. A
negative result would push toward other interpretations, such as
the presence of red-shifted Fe lines, that would be, conversely,
of the highest interest for the EOS of Neutron Stars.

\begin{figure}[ht]
\begin{center}
\epsfig{file=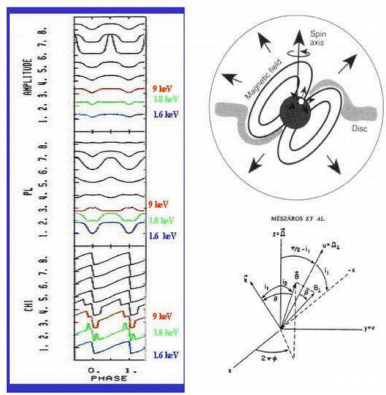, width=7cm}
\end{center}

\caption{The luminosity, polarization amount and polarization
angle for an X-ray pulsator at different energies and phases give
a direct measurement of the major geometric parameters.}
\label{polpul}
\end{figure}

\textbf{Accreting Magnetic Neutron Stars} are probably the best
known objects of X-ray Astronomy. But, while the mass can be
derived from the dynamics of the binary system, the rotation
period is derived from the pulsation and the magnetic field can be
measured with the cyclotron resonance energy, the mechanism of
transfer of plasma to the poles of the star is still derived
through complex models that fit the pulses shape and spectra.
These leave the geometric parameters (inclination of the magnetic
field versus the rotation, projection of the rotation axis on the
sky) as free parameters to be determined with the fit. As
demonstrated by (\cite{mesz88}), in such a system polarization is
modulated and the swing of the polarization angle with phase
directly measure the orientation of the rotation axis on the sky
and the inclination of the magnetic field: in Fig~\ref{polpul} the
case $45^{\circ}$, $45^{\circ}$. The same data will also identify
the emission model. In a fan beam the polarization will be
correlated in phase with the luminosity, while in a pencil beam it
will be anti-correlated.

\textbf{Millisecond accreting pulsar} represent a case of
particular interest. Millisecond coherent oscillation were
discovered in a number of low-mass X-ray binaries. The spectrum is
represented by the sum of a black-body like emission and a
Comptonized tail. If the Comptonization is produced in the
radiative shock at the antipodal accretion sites close to the
neutron star surface, we expect (\cite{viir04}) that the pulse
profile of the scattered radiation be phase-depedent polarized.
The amplitude of polarization degree and angular phase depend on
the geometry but also on the neutron star compactness. However
Flux and polarization will always be in phase. If instead is the
disk that intercept the flux from the antipodal spot
(\cite{sasSAX}) Doppler boost will make polarized flux trail or
lead the primary pulse of the direct emission. In both cases
polarimetry would probe the regions close to the surface of the
neutron star and complement timing and spectroscopy to the study
of the Equation of State of Neutron Stars.

\subsection{Matter under extreme gravitational fields }
\label{sec:ex} 
\begin{figure}[ht]
\begin{center}
\epsfig{file=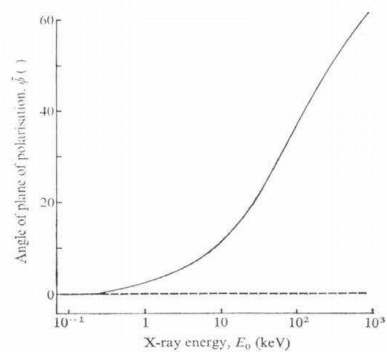, width=7cm}
\end{center}

\caption{The energy--dependent rotation of the polarization angle
for Cyg X-1 due to General Relativity effects (Stark \& Connors
1977)} \label{cygx1}
\end{figure}

Polarization from celestial sources may derive from
photons emitted in the innermost regions of an accretion disc
around a neutron star or a Black Hole which have, for symmetry
reasons, a polarization vector which, in the matter reference
frame, is either parallel or perpendicular to the disc axis. To
the distant observer, however, the polarization angle appears
rotated due to the parallel transport of the polarization vector
along the geodesics. This effect is significant only in the strong
gravity field regime, and the rotation is of course larger the
closer to the Black Hole the photon is emitted. In Galactic Black
Hole systems in the so--called 'soft states', X-rays may be
dominated by thermal emission from the disc, which may be
polarized due to Thomson scattering in a hot surface layer. The
decrease of the disc temperature with radius implies that higher
energy photons suffers a larger rotation than softer photons,
implying an energy dependence of the rotation of the polarization
angle. This very clear phenomenology  was predicted several years
ago (\cite{starcon77}, \cite{conn80}). The predicted rotation with
energy is shown in Fig~\ref{cygx1}, but has never been verified
due to the lack of the proper polarimeter.
\begin{figure}[ht]
\begin{center}
\epsfig{file=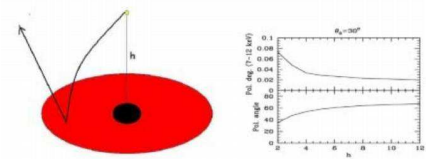, width=6cm}
\end{center}

\caption {The degree and angle of polarization of the reflected
radiation (right panel) as a function of the height of the primary
source, assumed for simplicity to be a point on the Black Hole
axis (left panel). Such a simple geometry seems able to explain
the otherwise puzzling variability of the relativistic iron line
in MCG-6-30-15 (Miniutti et al. 2003). \label{miniuttimatt}}
\end{figure}

In Active Galactic Nuclei, where thermal emission peaks in the UV
band, the disc reflects (and polarizes) via Compton scattering
part of the primary X-ray emission. The primary emitting region is
likely variable, with consequent variations of the regions of the
disc most illuminated. As a result, a time variation of the
polarization angle is expected. The gravitational bending has been
claimed to explain the different variability of relativistic
broadened line and continuum in MCG-6-30-15 (\cite{miniutti03}),
Variability of polarization angle with time can proof or disproof
this model (\cite{dovkarm}) or, more in general, any scenario in
which reflection originates in the innermost regions of disc, and
complement data from spectroscopy of relativistic lines. The
results in Fig.~\ref{miniuttimatt} have been obtained assuming
unpolarized primary emission, a rather conservative assumption.
Because the primary emission is likely due to Comptonization, it
is also expected to be significantly polarized (\cite{hama93},
\cite{poutvilh}); the net polarization degree is then likely to be
larger, still maintining time variations of the polarization
angle, as primary flux variations are likely related to variations
in the geometry of the emitting region.

Whatever the details, it is hard to imagine other effects
producing energy and/or time variations of the polarization
angles. Thence, if observed, the rotation would provide a strong
signature of General Relativity effects, and a test of this theory
in regimes that can not be probed within the solar system.

\subsection{Cosmic Accelerators }
\label{sec:ex} Acceleration of particles is the clue of Cosmic Ray
Physics. Super Nova Remnants (both plerionic and shell-like) are
the best candidates for acceleration of electrons. They emit up to
TeV energies. Jets in AGNs and in Galactic Sources GRBs and their
Afterglows are for sure the site of acceleration but the structure
and origin of the magnetic field and the structure and energy
distribution of jets are still open questions.
\subsubsection{Cosmic Accelerators by themselves} 
\begin{figure}[ht]
\begin{center}
\epsfig{file=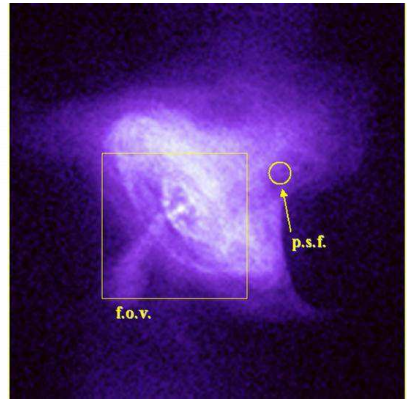, width=7cm}
\end{center}

\caption{Crab Nebula and pulsar as imaged by Chandra, superimposed
the f.o.v. and the p.s.f. of XPOL.\label{crabW}}
\end{figure}

The prototype of
all cosmic accelerators, the Crab, is also the only source for
which X-ray polarization has been actually measured. In 1978 OSO-8
 satellite measured a polarization of 19.2 $\pm$
1.0 $\%$ (\cite{weis78}). But this data only refer to the average
properties of the system. In Fig.~\ref{crabW} we show the image of
Crab from Chandra (\cite{weis00}), showing that several well
distinguished subsystems contribute to the emission,including an
inner torus, an outer torus, the pulsar and two polar jets. We
also show the f.o.v. of XPOL and the size of a pixel: XPOL can
resolve each of these structures and measure the coupling of the
magnetic field with accelerated electrons.
A major contribution to understanding the acceleration processes
in SNR is coming from spectra. The presence of a hard component
(extending up to TeV) provides the evidence for acceleration even
in shell like remnants. These measurements could be complemented
with X-ray Polarimetry, providing, with a much better angular
resolution, the degree of order of the particles and the role of
turbulence. The angular resolved measurement of the polarization
angle provides an information on the geometry of the shocked site,
which cannot be derived from spectra.

\subsubsection{Cosmic Accelerators to test Quantum Gravity} Quantum
Gravity should be effective on the Planck Energy scale
($E_{QG}$=10$^{19}$ GeV). But the hypothesized existence of
space-time foam can produce detectable effects on radiation
propagating on very long distance scale.  One of the major
approach to quantization of Gravity is the Loop QG that predicts
birefringence effects. The result is a difference of light
velocity for the two states of circular polarization:

\begin{equation}
 v_{+}=c[1+\chi(E/E_{QG})^{n}]
\end{equation}
\begin{equation}
v_{-}=c[1-\chi(E/E_{QG})^{n}]
\end{equation}

Since linear polarization is a superposition of the two
eigenstates of circular polarization, the plane of linear
polarization is subject to a rotation along the path. This
birefringent behavior is energy dependent. The index \textit{n} is
different in different models.  The linear dependence of the
velocity on ($E$/$E_{QG}$) (n=1) is already ruled out by the 1976
measurement of X-ray polarization of Crab, already mentioned as
the only positive result of X-ray polarization so far
(\cite{kaaret03}).

But the quadratic case is still viable. Following
 \cite{mitrofanov03}, the expected rotation is:
\begin{equation}
\Delta \phi \simeq \chi \cdot(D/hc) \cdot E^{3}/E_{QG}^{2} \simeq
10^{-22} \cdot \chi \cdot(E/10keV)^{3} \cdot D
\end{equation}
It requires very large distances to be tested . The effect is
larger at higher energies. In high energy $\gamma$-rays
polarimetry is in principle viable, but is not foreseen in
practice. Scattering polarimetry in soft $\gamma$-rays is
possible, but requires high fluxes and high polarizations for a
detection. X-rays are, therefore, the highest energy band where
sensitive polarimetry of remote sources can be performed. If a
source is emitting mainly for synchrotron the polarization angle
should be the same at different energies. Any class of sources
dominated by synchrotron can be a good candidate. Of course, to
confirm that this study is feasible, objects of the selected class
should have been deeply tested and their phenomenology well
understood.

Gamma Ray Burst afterglows are a good candidate, because can be
studied up to Z=4, but we do not really know how polarized they
are (a few $\%$ in the optical emission) and the polarization
could be variable on short timescales. The observation of
afterglows should be performed in an early phase, before they are
too faint to be measured. This requires a dedicated strategy of
satellite operations. Blazars are likely a good alternative. Even
though blazars that can be arrived are closer they should be
highly polarized and the dependence on energy should be moderate
or null in the X-Rays, as far as we observe the part of the
spectrum where synchrotron is largely dominant, or where SSC is
well established (\cite{pout}, \cite{celmatt}). The joint
variability at different energies, in flaring episodes, suggests
that photons come from the same region and, for dominant
synchrotron, could have the same polarization angle. So the method
could be extended to a larger energy band. QSOs are less
attractive, because of the above mentioned effects of
gravitational bending. Seyfert-2 galaxies could, conversely, be a
candidate class of objects.
\section{A polarimeter for a NGXT}
Some of the mentioned phenomena could be studied with a
photoelectric polarimeter in the focus of a large area telescope,
foreseen for a New Generation X-ray Telescope in the frame of the
ESA Cosmic Vision 2015-2025. We call this instrument
\textbf{XPOL}. It is based on two new, but existing, technologies:
the GEM and the VLSI read-out. The baseline filling would be a
50\% mixture of Ne and DME at 1 Atm pressure. We did all the
estimations of sensitivity with this mixture, but there is a
margin of improvement, with mixtures and pressures, better tuned
with the band-pass of the telescope. The other components (the Be
window, the gas cell, the field forming rings) are established
technology for proportional counters. The read-out electronics
would be relatively simple, as the ASIC chip already performs a
pre-selection of the window around the track. Data of faint
sources can be directly downloaded. Data from the brightest
sources should be recorded onboard, preprocessed and downloaded in
a compressed form. Including  HVPS, control electronics and a
buffle, to exclude the X-ray sky, the needed resources are of the
order of 10 kg and 15 watts.

\begin{figure}[ht]
\begin{center}
\epsfig{file=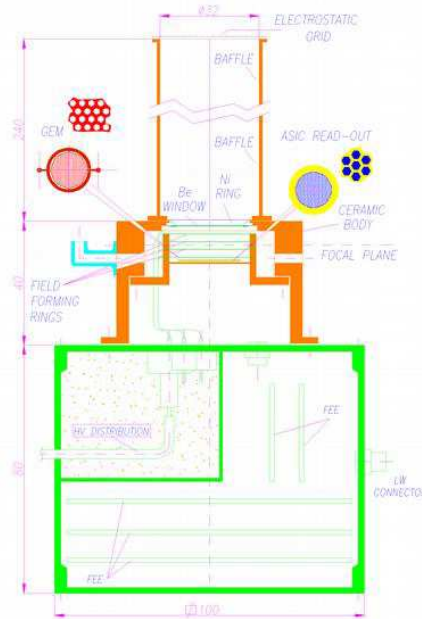, width=9cm}
\end{center}

\caption{Drawing of XPOL \label{xpol}}
\end{figure}

In Fig.~\ref{xpol} we show the drawing of a detector based on
these concepts.

\begin{figure*}[ht]
\begin{center}
\epsfig{file=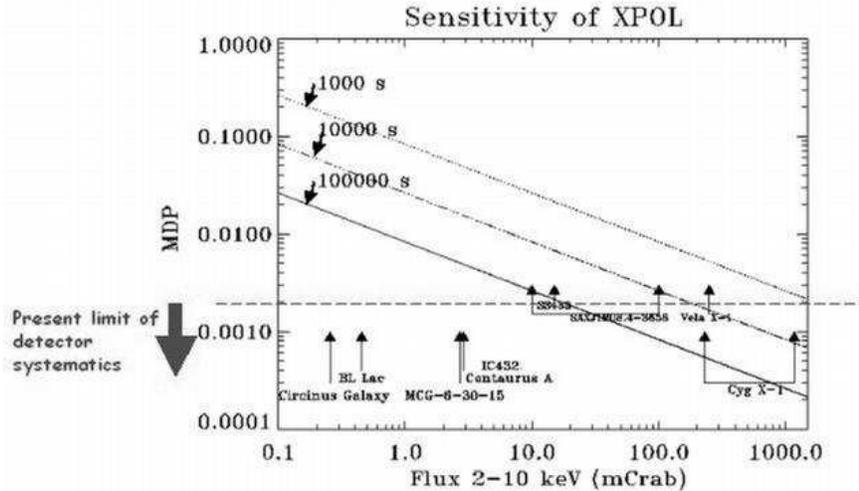, width=13cm}
\end{center}

\caption{Minimum Polarization Detectable by XPOL on a few
representative sources \label{sens}}
\end{figure*}

In Fig.~\ref{sens} we show the \textbf{sensitivity} of XPOL for
different exposure times and for a few representative sources.
With observations of one day we can measure the polarization of
several AGNs down to \% level. Most of objectives outlined in this
paper can be achieved with shorter pointing. The impact of
systematics shown in the figure refers to the detector. With
bright galactic sources this limit is arrived in $10^{3}$ s.
Therefore, for these sources, phase or energy resolved polarimetry
can be performed in a few hours.

\section{Conclusions}
\begin{itemize}
\item  X-ray Polarimetry complements other techniques for deep physics of extreme objects
\item  X-ray Polarimetry is now feasible
\item  X-ray Polarimetry needs a large optics but only small resources in a shared focal plane
\item  X-ray Polarimetry is worthwhile and should be part of a Cosmic Vision 2020
\end{itemize}

\begin{acknowledgements}

The development of Micropattern Gas Chamber has been supported by
INFN, INAF and ASI. It also benefitted of a fraction of Descartes
Prize 2002 of the European Commission.

\end{acknowledgements}

\end{document}